\pdfoutput=1
\documentclass[sigplan,nonacm]{acmart}

\usepackage{kim-paper-macros}

\hypersetup{pdfsource={}}

\newcommand{\sys}{{\textsf{Leto}}\xspace}

\newcommand{\jitcpkt}{\textsf{JIT-Ckpt}\xspace}
\newcommand{\gemini}{\textsf{Gemini}\xspace}
\newcommand{\moevement}{\textsf{MoEvement}\xspace}

\newcommand{\geminiplus}{\textsf{Gemini+}\xspace}
\newcommand{\moevementplus}{\textsf{MoEvement+}\xspace}
\newcommand{\jitcpktplus}{\textsf{JIT-Ckpt+}\xspace}

\newcommand{\pythonloc}{7.8K\xspace}
\newcommand{\cpploc}{2.4K\xspace}

\newcommand{\qwen}{Qwen\xspace}
\newcommand{\olmoe}{OLMoE\xspace}
\newcommand{\dsmoe}{DeepSeek-MoE\xspace}
\newcommand{\gptmoe}{GPT-MoE\xspace}
\newcommand{\kimimoe}{Kimi-MoE\xspace}
\newcommand{\dsvtwo}{DeepSeek-V2\xspace}
\newcommand{\dsvthree}{DeepSeek-V3\xspace}
\newcommand{\dsvfour}{DeepSeek-V4\xspace}

\newcommand{\hof}{HOF\xspace}
\newcommand{\ff}{FF\xspace}
\newcommand{\hofs}{HOFs\xspace}
\newcommand{\ffs}{FFs\xspace}
\newcommand{\rmp}{RMP\xspace}
\newcommand{\rmu}{RMU\xspace}

\renewcommand\footnotetextcopyrightpermission[1]{}
\begin{document}

\title{\textsf{Leto}: Fast In-Place Recovery for LLM Training on Surviving Hardware}

\author{Geon-Woo Kim}
\affiliation{%
  \institution{The University of Texas at Austin}
  \city{Austin}
  \state{Texas}
  \country{USA}
}
\email{gwkim@utexas.edu}

\author{Joon Ha Kim}
\affiliation{%
  \institution{The University of Texas at Austin}
  \city{Austin}
  \state{Texas}
  \country{USA}
}
\email{jhk.james1110@utexas.edu}

\author{Daehyeok Kim}
\affiliation{%
  \institution{The University of Texas at Austin}
  \city{Austin}
  \state{Texas}
  \country{USA}
}
\email{daehyeok@utexas.edu}

\begin{abstract}
Hardware-operable failures (\hofs) interrupt large language model (LLM) training but permit recovery on the same hardware without reset, repair, or replacement.
Existing recovery systems nevertheless reload checkpoints, recompute lost progress, and rebuild process state, idling GPUs that could otherwise continue training.

We present \sys, a fault-tolerant training system that leverages surviving hardware to enable efficient in-place recovery.
Our key insight is that the state needed to resume training can be retained or prepared outside the active training process while remaining on the same hardware.
\sys retains the working model state and the reusable process state, and preinitializes the remaining state in a shadow trainer.
We devise two-tier erasure protection and chunk-level transactional updates to keep the retained model state recoverable and consistent, and reclaim the shadow state when active training needs its GPU memory.
Evaluation on 6- and 72-GPU NVIDIA A100 clusters shows that \sys recovers 3.6--6.5$\times$ faster than the best-performing checkpointing baselines and improves productive training time by up to 13.7 percentage points.
Large-scale simulation shows over 95\% productive training time on a 131,072-GPU cluster.
\end{abstract}

\maketitle

\section{Introduction}
\label{sec:intro}

Training large language models (LLMs) at scale can occupy thousands of GPUs for weeks or months~\cite{llama3,deepseek_v3}.
A single failed training process, or \emph{trainer}, can stall the entire job and idle otherwise healthy GPUs.
We focus on \emph{hardware-operable failures} (\hofs), in which trainers fail but training can resume on the same nodes and GPUs without hardware reset, repair, or replacement.
Such failures can result from contained memory errors or temporary network faults that disrupt a trainer while leaving its hardware usable~\cite{nvidia_gpu_memory_error_mgmt,robustinfra}.
Restarting training on the same hardware resolved 22.7\% of failures across 19 large-scale production jobs~\cite{robustinfra}, while a 256-GPU study reports that process restarts can resolve 73\% of failures~\cite{unicron}.
These reports show that HOFs account for a substantial share of training failures.

Today's recovery systems for LLM training do not fully exploit this surviving hardware.
They typically restore \emph{model state}, including parameters and optimizer state, from the latest checkpoint and recompute the work completed since that checkpoint~\cite{moevement,gemini,jitckpt,bytecheckpoint,pccheck,checkfreq}.
They also rebuild \emph{process state}, such as CUDA contexts and communication resources, for the trainer that takes over after a failure.
Keeping checkpoints in memory speeds up loading, while preinitializing trainers on spare nodes reduces process setup time~\cite{trainmover}.
Yet recovery still loads an older model state and repeats computation to regain pre-failure progress; the spare must also finish process setup that depends on which trainer it replaces.
In our measurements, recovering \kimimoe~\cite{kimilinear} on 64 NVIDIA A100 GPUs takes 88 seconds with both optimizations, including 35 seconds for remaining initialization and checkpoint loading and 53 seconds for recomputation (\autoref{subsec:cost}).

This paper asks: \emph{Can a replacement trainer reuse state on surviving hardware to minimize recovery overhead?}
Our observation is that, for a \hof, the state needed to resume training can be retained or prepared outside the active trainer while remaining on the same hardware.
The working model state and the reusable process state could remain accessible to a replacement process, avoiding checkpoint loading and recomputation of completed iterations.
For process state that cannot outlive its owner, the unchanged hardware and training rank offer a second opportunity to initialize that state in another process before a failure.
Together, state retention and preinitialization could remove much of the work that makes recovery expensive.

We present \sys, a fault-tolerant training system that realizes these opportunities through \emph{in-place state retention} and \emph{in-place preinitialization}.
\sys pairs each active trainer with a lightweight \emph{state manager} and a \emph{shadow trainer} on the same node.
The state manager owns the memory holding the working model state and the reusable process state, allowing their allocations to survive the active trainer.
The shadow trainer prepares the remaining process state, including CUDA contexts and NCCL communicators, for the same GPUs and training rank.
After a \hof, the shadow takes over and reattaches to the retained state.

Realizing this approach correctly and with low overhead raises three challenges.
First, retained state remains exposed to memory faults that can make multiple pages inaccessible even while the GPU remains operable.
Such memory faults are a significant source of training and infrastructure failures~\cite{llama3,superbench}.
Second, a failure can interrupt an in-place model update, leaving retained state partially updated and inconsistent across trainers.
Third, preinitializing a shadow consumes GPU memory alongside active training, whose changing memory demand can make the shadow cause an out-of-memory (OOM) failure.

We address these challenges through three key ideas.
First, GPU-resident model state can be protected by redundancy in host memory, provided that the redundancy is itself protected.
We therefore devise two-tier erasure protection in a \emph{Resilient Memory Pool}, avoiding dedicated GPU memory for persistent parity.
Second, fine-grained optimizer updates and incremental parity updates make it possible to preserve progress at chunk boundaries.
We exploit this structure in the \emph{Resumable Model Updater} to recover interrupted updates without full-model snapshots.
Third, preinitialization can be deferred without losing training progress.
Our \emph{preinitialization controller} treats this state as reclaimable, trading recovery-time initialization for memory headroom during active training.

We implement \sys{} on Megatron-Core, PyTorch, and NCCL in approximately \pythonloc{} lines of Python and \cpploc{} lines of C++/CUDA, integrating shared state allocation, shadow reclamation on allocation failure, and GPU parity kernels.
We will open-source our implementation.

We evaluate one dense and four mixture-of-experts (MoE) LLMs on two NVIDIA A100 clusters, the larger with nine eight-GPU nodes (72 GPUs).
With one \hof injected every 10 minutes, \sys recovers 3.6--6.5$\times$ faster than the best-performing state-of-the-art checkpointing baselines across measured workloads, including variants with spare-based preinitialization, improving productive training time by up to 13.7 percentage points.
Large-scale simulation further shows that \sys sustains over 95\% of training time in productive computation on a 131,072-GPU cluster, even with a 10-minute mean time between \hofs.

\section{Background and Motivation}\label{sec:background}

\begin{figure}[t]
  \centering
  \includegraphics[width=\columnwidth]{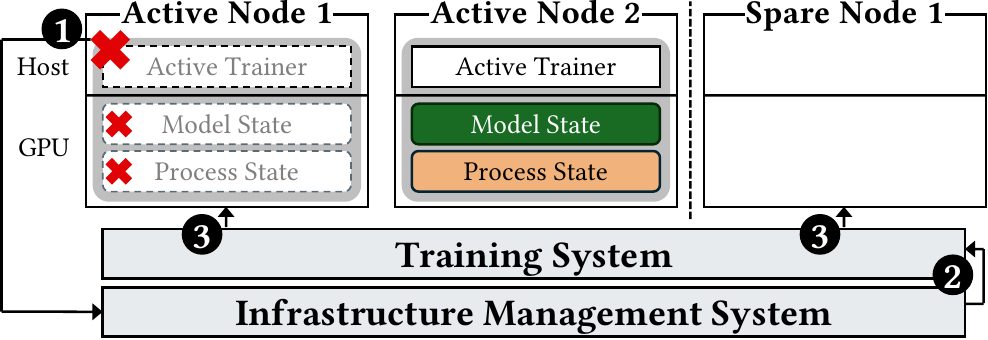}
  \caption{Training-job recovery. The infrastructure \circledblack{1} detects a trainer failure and \circledblack{2} checks hardware operability, then \circledblack{3} triggers in-place recovery (\hofs only) or spare-based recovery (either failure class).}
  \Description{}
  \label{fig:failure_handling_by_training_infra}
\end{figure}

\subsection{Hardware-Operable Failures}
\label{sec:hof}
A hardware fault can interrupt a training job without making its hardware unusable.
We call a job failure a \textit{hardware-operable failure} (\hof) if one or more training processes terminate or hang, but training can resume on the same nodes and GPUs without hardware reset, repair, or replacement.
A \textit{fatal failure} (\ff), in contrast, requires such remediation before the affected hardware can be reused.
Both interrupt training; the distinction is whether the affected hardware can be reused without remediation.

Modern hardware can contain faults within a process or memory region while leaving the device operable~\cite{ocp_gpu_ras_2025}.
For example, a contained uncorrectable error in GPU high-bandwidth memory (HBM) can terminate the affected process while leaving the GPU available to run a replacement without a reset~\cite{nvidia_gpu_memory_error_mgmt}.
The faulty memory pages may become inaccessible, but the remaining memory is still usable.
Similarly, a temporary network link outage can leave training processes hung in collective communication after connectivity returns, requiring a process restart while the nodes remain usable~\cite{robustinfra,revisiting,reccl}.

Production reports show that many failures can be resolved by restarting training on the same hardware.
Across 19 large-scale training jobs at ByteDance, 22.7\% of failures were resolved by restarting training on the same hardware~\cite{robustinfra}.
Separately, a 256-GPU study reports that 73\% of failures can be resolved by restarting the training processes~\cite{unicron}.

We expect \hofs to become more prevalent for two reasons.
First, growing GPU counts and memory capacities increase fault exposure~\cite{nvidia_nvlink_roadmap,nvidia_rubin}.
A production study reports 3.2$\times$ more frequent uncorrectable memory errors per GPU on newer H100 GPUs (96\,GB of HBM) than on A100 GPUs (40\,GB), potentially reflecting the increase in memory capacity~\cite{cui2025story}.
Second, improved fault containment lets more errors interrupt processes without disabling hardware, enabling recovery on the same hardware~\cite{nvidia_gpu_memory_error_mgmt,ocp_gpu_ras_2025}.

\subsection{Failure Recovery: Where and What to Recover}
When a training process, or \textit{trainer}, fails, another trainer must take over its role in the job.
The infrastructure decides where this \textit{replacement trainer} will run~\cite{robustinfra,llama3,megascale,revisiting,acme}.
As \autoref{fig:failure_handling_by_training_infra} illustrates, it \circledblack{1} detects the failure, \circledblack{2} identifies the affected nodes and checks whether their hardware remains operable, and \circledblack{3} triggers recovery~\cite{robustinfra,deng2025minder,superbench}.

\mypara{Where to recover.}
Recovery follows one of two placement policies.
\textit{In-place recovery} launches replacement trainers on the same nodes and GPUs~\cite{torchrun,nvrx}.
\textit{Spare-based recovery} launches them on spare nodes that have already passed hardware health checks~\cite{robustinfra,trainmover}.
An \ff requires spare-based recovery to resume training without waiting for hardware remediation.
A \hof permits either form of recovery, with in-place recovery often favored to keep spare nodes available for other failures~\cite{robustinfra,unicron,nvrx}.

\mypara{What to recover.}
Regardless of where it runs, a replacement trainer needs two classes of state before it can resume training.
\textit{Model state} includes the model parameters and optimizer state updated over successive training iterations.
\textit{Process state} includes the runtime resources needed to execute those iterations, such as CUDA contexts, communication resources, and GPU and host memory buffers.

These states normally belong to the trainer process, so launching a new process on the same GPU does not automatically restore them.
Recovery systems such as \gemini~\cite{gemini} and \moevement~\cite{moevement} restore the model state and initialize replacement trainers even after a \hof.
Thus, \emph{recovering on the same hardware does not by itself avoid rebuilding the model and process state needed to resume training}.

\subsection{The Cost of Reconstructing State}
\label{subsec:cost}

\begin{figure}[t]
  \centering
  \begin{subfigure}{\columnwidth}
    \centering
    \includegraphics[width=\columnwidth]{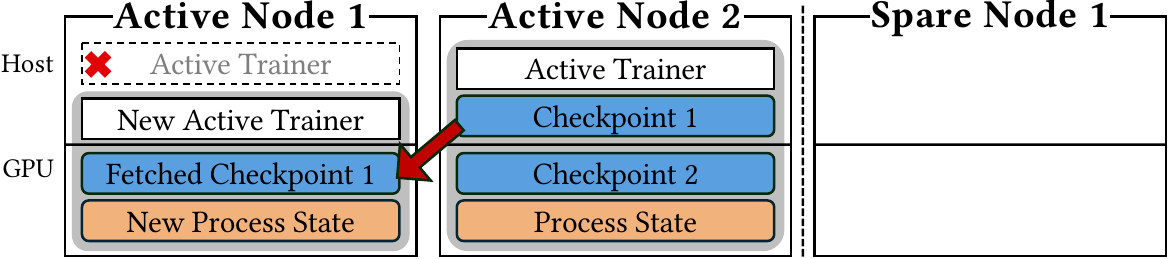}
    \caption{In-place recovery.}
    \label{fig:failure_handling_by_inmem}
  \end{subfigure}
  \begin{subfigure}{\columnwidth}
    \centering
    \includegraphics[width=\columnwidth]{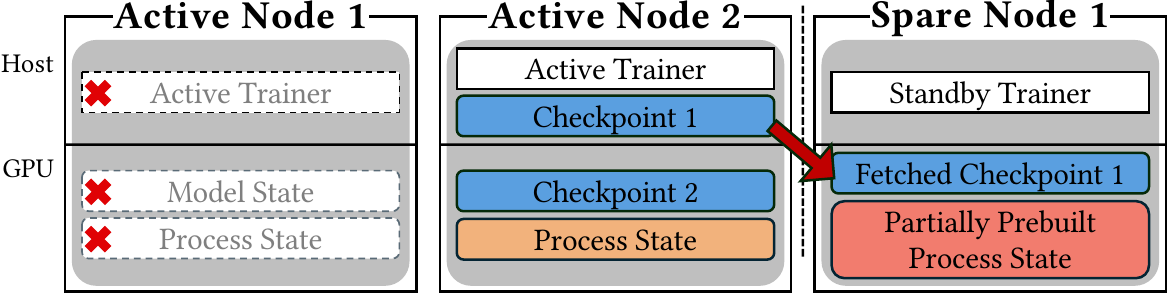}
    \caption{Spare-based recovery with preinitialization.}
    \label{fig:failure_handling_by_inmem_preinit}
  \end{subfigure}
  \caption{\hof recovery using in-memory checkpointing (\eg Gemini~\cite{gemini}), with and without spare-based preinitialization.}
  \Description{}
  \label{fig:failure_handling_by_sota}
\end{figure}

Most training systems today rely on checkpointing to restore model state after a failure~\cite{moevement,gemini,jitckpt,bytecheckpoint,pccheck,checkfreq}.
Spare-based preinitialization complements checkpointing by preparing some process state in advance, but a preinitialized trainer still needs the model state before it can resume training.
\autoref{fig:failure_handling_by_sota} illustrates these recovery paths.

\mypara{Checkpointing.}
Training systems periodically save copies of model state, called \textit{checkpoints}, in storage that survives a trainer failure, such as remote host memory~\cite{gemini,moevement} or local disk~\cite{checkfreq,pccheck}.
Under in-place recovery (\autoref{fig:failure_handling_by_inmem}), a replacement trainer starts on Active Node~1, initializes its process state, and loads its latest checkpoint from Active Node~2.
The trainers roll back to their checkpoints and recompute the work completed since those checkpoints.
Keeping checkpoints in memory speeds up loading, but recovery still restores an older model state and repeats computation to regain pre-failure progress.

\mypara{Spare-based preinitialization.}
A standby trainer on a spare node can initialize some process state before an unexpected failure, as in TrainMover~\cite{trainmover}.
In \autoref{fig:failure_handling_by_inmem_preinit}, the standby on Spare Node~1 already holds resources such as pinned host memory buffers.
However, which trainer it replaces is determined only after a failure.
The standby must then finish setup for the assigned rank in the distributed job, including connections to the other trainers.
For checkpoint-based recovery, the standby then loads the failed trainer's checkpoint, and the job recomputes lost work.
Preinitialization reduces process-state setup time while leaving checkpoint loading and recomputation necessary.

\mypara{Recovery cost.}
We measure this combination by training \kimimoe on 64 NVIDIA A100 GPUs with \geminiplus, which adds spare-based preinitialization to \gemini's in-memory checkpointing.
Recovery takes 88 seconds: 35 seconds for remaining initialization and checkpoint loading, and 53 seconds to recompute lost progress (\autoref{subsec:perf_breakdown}).
The time for initialization and checkpoint loading falls from 59 seconds with \gemini to 35 seconds with \geminiplus.
Even with both techniques, recovery spends tens of seconds preparing replacement trainers and restoring pre-failure training progress.

Recovery latency reduces productive training time across the distributed job.
The Effective Training Time Ratio (ETTR) measures the fraction of wall-clock time spent on productive training~\cite{revisiting}.
Prior work projects a mean time to failure (MTTF) of 0.23 hours for 131,072-GPU jobs~\cite{revisiting}.
At that projected failure rate, an illustrative recovery cost of 88 seconds per failure would consume roughly 10.6\% of wall-clock time, before accounting for other overheads.

\subsection{Opportunities for In-Place Recovery}
\label{subsec:opp}

These recovery paths rebuild state and replay training even when the affected hardware remains operable, limiting productive training time.
\textbf{\em Can the surviving hardware help avoid this process?}
It offers two complementary opportunities for in-place recovery.

\mypara{Opportunity 1: Retain state in place.}
State that can outlive the training process could remain resident on the still-operable hardware (\eg GPU HBM or host memory) and be accessed directly by a recovering trainer.
Retaining the working model state and the reusable process state (\eg GPU communication buffers) outside the trainer's failure domain could therefore avoid checkpoint loading, recomputation of completed iterations, and buffer recreation.

\mypara{Opportunity 2: Preinitialize process state.}
State tied to the training process, such as CUDA contexts and NCCL communicators, must be created for the recovering trainer.
However, its initialization depends on the GPUs, network interfaces, and rank that the trainer will use.
By preserving these assignments, in-place recovery could enable this state to be initialized before a failure, moving the remaining process-state initialization off the recovery critical path.

Together, these opportunities could shorten in-place recovery through state reuse and advance preparation.

\section{\sys Overview}\label{sec:overview}

We present \sys, a fault-tolerant training system that realizes both opportunities in \autoref{subsec:opp} to recover from \hofs in place.
\sys pairs each active trainer with a lightweight \emph{state manager} and a \emph{shadow trainer}, both on the same node.
These components enable two recovery mechanisms:

\noindent\textbf{In-place state retention.}
The state manager retains the working copy of the model state and the reusable process state (\eg GPU communication buffers) outside the active trainer's failure domain.
After a \hof, the replacement trainer accesses this state directly, avoiding checkpoint loading and recomputation of completed iterations (\autoref{fig:failure_handling_by_inmem}).

\noindent\textbf{In-place preinitialization.}
The shadow trainer initializes the remaining process state, including CUDA contexts and NCCL communicators, before a failure.
Because it will take over the active trainer's GPUs, network interfaces, and rank, it can prepare states that are specific to these assignments in advance, unlike a spare awaiting its replacement assignment (\autoref{fig:failure_handling_by_inmem_preinit}).
Upon a \hof, it takes over the active role, reusing the state prepared in advance.

\subsection{Design Goals}\label{subsec:goals}
Faster recovery must preserve training results and improve effective throughput.
We therefore pursue three design goals:

\mypara{G1: Guarantee exact recovery.}
Given the same input-batch and random-number-generator sequences, initial model state, and training configuration, and assuming deterministic execution, the model state and loss at each completed iteration must match those of a fault-free execution, regardless of failure timing.
Small perturbations can silently steer training to a different optimum and are hard to diagnose~\cite{ma2025understandingsdc,jiang2025training}.

\mypara{G2: Minimize steady-state overhead.}
\sys should minimize iteration-time overhead to improve effective throughput even when \hofs are a minority of failures.

\mypara{G3: Minimize GPU memory overhead.}
\sys should minimize the GPU memory consumed by recovery support, including prebuilt process state, to accommodate LLM training workloads whose HBM utilization often exceeds 90\%~\cite{trainmover}.

\subsection{Challenges}\label{subsec:challenges}
Realizing state retention and shadow preinitialization under these goals raises three challenges.

\mypara{C1: Memory page failures in retained state.}
The state manager retains state outside the active trainer's failure domain, but that state remains exposed to faults in the underlying memory.
A contained HBM error can leave the GPU operable while making multiple pages of retained state inaccessible~\cite{nvidia_gpu_memory_error_mgmt,amd_umc_v12}.
\sys must tolerate multiple page failures with minimal GPU-memory overhead.

\mypara{C2: Inconsistent model state from mid-update failures.}
The state manager retains the working copy that the active trainer updates in place.
A failure during an update can leave this copy partially updated and inconsistent across ranks, even when all its pages survive.
\sys must recover a consistent training state without the memory cost of a full extra copy or the steady-state cost of snapshotting.

\mypara{C3: Memory contention from prebuilt process state.}
The shadow trainer's preinitialized GPU state shares HBM with active training, whose memory demand varies over time, especially with MoE routing imbalance~\cite{zhou2022expertchoice,megatron_moe_docs}.
A shadow that initially fits can therefore later cause an out-of-memory (OOM) failure in the active trainer.
\sys must preserve preinitialization's recovery benefit while meeting active training's GPU-memory needs.

\subsection{Key Ideas}
\label{sec:ideas}
We propose three techniques to address these challenges.

\mypara{I1: Resilient memory pool with two-tier erasure protection~(\autoref{sec:rmp}).}
We design the \emph{Resilient Memory Pool} (RMP) to retain model and reusable process state, protecting model contents against memory page failures.
RMP uses erasure-code parity to reconstruct lost contents, but storing that parity in HBM would consume scarce training memory.
We therefore devise \emph{two-tier erasure protection} that keeps both persistent parity tiers in host memory.
The first tier protects retained HBM data, while the second protects first-tier parity against host-memory failures.

\mypara{I2: Resumable model updater with chunk-level transactions~(\autoref{sec:rmu}).}
We design the \emph{Resumable Model Updater} (RMU) to make in-place model updates crash-consistent.
We exploit the fine granularity of model updates and incremental erasure-code parity updates to devise chunk-level transactions that commit the model state and its parity together.
This lets each rank resume its interrupted update and reach a consistent iteration boundary without snapshotting the full model.
RMU also reuses activation memory for update buffers once backward propagation completes, avoiding any increase in peak HBM usage.

\mypara{I3: OOM-free preinitialization with runtime reclamation~(\autoref{sec:mpreinit}).}
We observe that the training runtime's memory allocator provides an opportunity to reclaim memory before reporting an OOM, and that prebuilt process state can be recreated if discarded.
We design a \emph{preinitialization controller} that uses this opportunity to reclaim the shadow trainer's GPU memory when active training needs it.
By adapting preinitialization to available memory, our design preserves fast recovery when memory permits while preventing preinitialization from causing training OOMs.

\begin{figure}[t]
  \centering
  \includegraphics[width=\columnwidth]{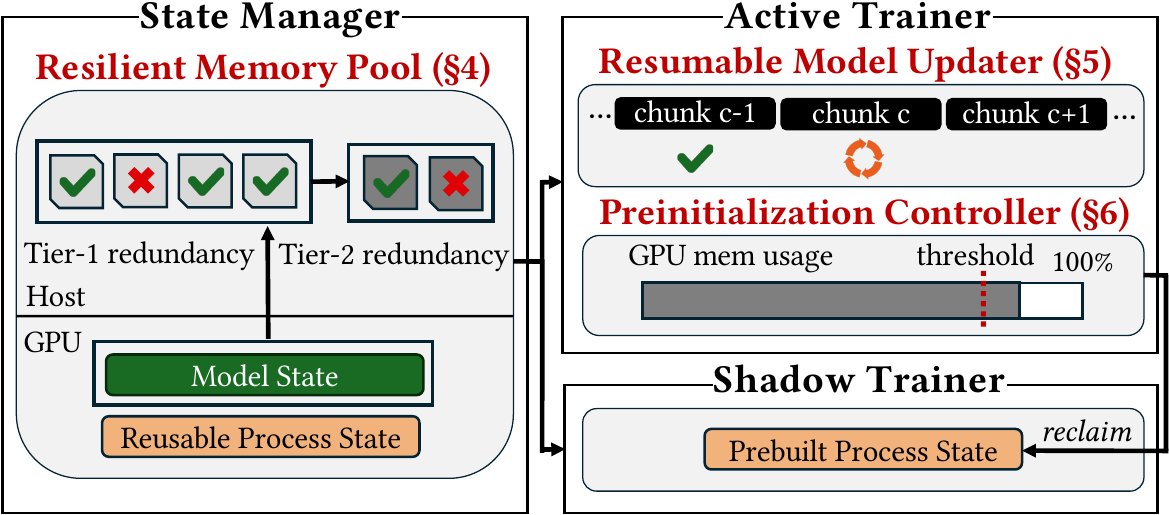}
  \caption{\sys's architecture, with a state manager for retained state and a shadow trainer for preinitialization.}
  \Description{}
  \label{fig:architecture}
\end{figure}

\subsection{\hof Recovery Workflow in \sys}\label{subsec:hof_recovery}
\autoref{fig:architecture} shows the components involved in \sys's \hof recovery.
When an active trainer fails, RMP retains its model state and reusable process state, and localizes and marks any pages made inaccessible by the failure.

To use their prebuilt NCCL communicators, the shadow trainers take over together.
A shadow whose GPU state was reclaimed completes its deferred initialization before the role switch.
Surviving active trainers release their process-owned GPU memory and become shadows; \sys launches a new shadow for the failed trainer.

Each promoted trainer reattaches to the retained state, and its \rmu reconstructs failed protected pages using the available data and parity.
If no rank has begun updating its model state, training restarts the interrupted iteration.
Otherwise, each rank's \rmu restores any interrupted chunk and resumes from its last committed chunk boundary until all ranks complete that iteration's update.

\sys targets \hof recovery; an \ff prevents reuse of the affected hardware and may destroy its retained state.
For these failures, \sys uses existing checkpointing systems unchanged and restores training from their checkpoints (\autoref{subsec:cost}).

\section{Resilient State Retention}\label{sec:rmp}

Retained state must remain recoverable despite memory page failures, with minimal GPU-memory overhead.
We design the \emph{Resilient Memory Pool} (RMP), a memory-management component within each state manager that owns allocations for retained model and reusable process state.

\subsection{Two-Tier Erasure Protection}
GPU HBM uses error-correcting codes (ECC) to correct single-bit errors but cannot repair all error patterns~\cite{nvidia_gpu_memory_error_mgmt}.
When a double-bit flip occurs within an ECC-protected domain (\eg 32 bytes in HBM3 and later), the memory controller detects the uncorrectable error and reports it to the GPU driver.
The driver identifies the coarse-grained failure domain containing the error (\eg a DRAM row) and retires all virtual memory pages mapped to that physical region (\eg 16 virtual pages on AMD MI300X~\cite{amd_umc_v12}), preventing further access to them~\cite{nvidia_gpu_memory_error_mgmt,amd_umc_v12}.
A localized fault thus creates multiple page \emph{erasures}, meaning that RMP can identify the failed pages but cannot read their contents.

We use Reed--Solomon (RS) erasure coding~\cite{reed_solomon} to reconstruct these lost pages.
RS computes redundant parity pages from data pages and stores them separately; the number of parity pages determines how many page erasures the code can tolerate.
Within the code's configured tolerance, surviving data and parity suffice to reconstruct the lost contents.

The challenge is to provide enough parity for multiple page failures without reducing the HBM available to memory-intensive LLM training~\cite{llama3,trainmover}.
For example, \kimimoe uses up to 14.1~GiB of GPU memory for its model state in our evaluation (\autoref{subsec:setup}).
With the minimum HBM virtual-memory allocation granularity of 2~MiB on NVIDIA A100 GPUs, tolerating 16 page failures within each 2~GiB DRAM die requires 16 parity pages per die.
Storing this parity in HBM would consume an additional 256~MiB for this model state.
Stronger protection raises this cost: tolerating two concurrent failures, each spanning 32 pages within a DRAM die, requires 1~GiB of additional HBM.

\begin{figure}[t]
  \centering
  \includegraphics[width=\columnwidth]{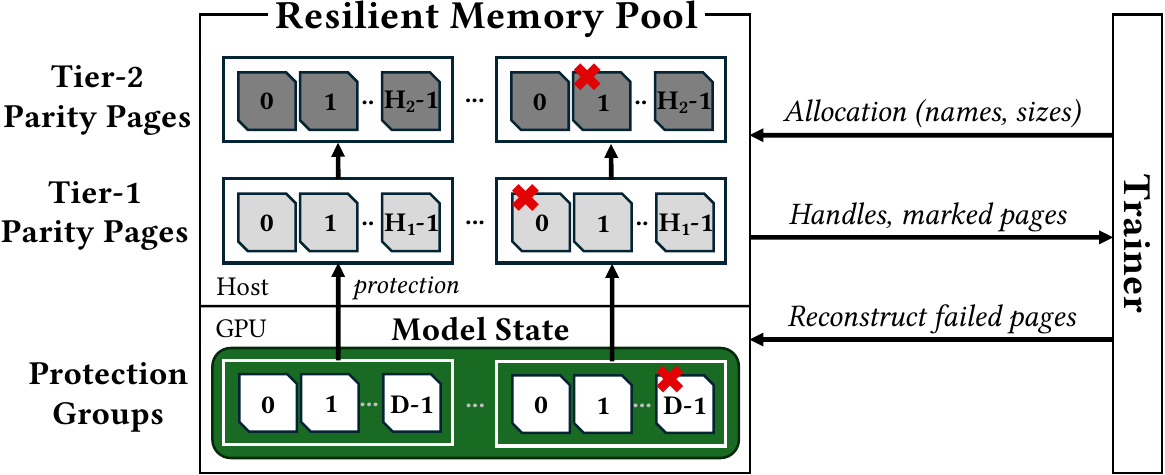}
  \caption{RMP organizes model state into protection groups of \(D\) HBM data pages, protected by \(H_1\) first-tier and \(H_2\) second-tier parity pages in host memory.}
  \Description{}
  \label{fig:rmp}
\end{figure}

\mypara{Our approach.} Our key insight is that retained data and their recovery redundancy need not occupy the same memory tier.
We exploit this separation to devise \emph{two-tier erasure protection}, which keeps persistent parity in host memory to avoid reserving HBM for both parity tiers.
As illustrated in \autoref{fig:rmp}, RMP partitions the retained model state into fixed-size \emph{protection groups} of HBM pages and stores each group's first-tier parity in host memory.
Host memory, however, can also suffer uncorrectable ECC errors, leaving that parity unavailable.
We therefore protect the first-tier parity with a second tier of parity pages, also in host memory.

Each protection group contains \(D\) data pages, with \(H_1\) first-tier and \(H_2\) second-tier parity pages; each page has size \(S\).
By default, RMP uses \(D=1024\), \(H_1=16\), \(H_2=1\), and \(S=2\)~MiB, yielding a 2~GiB protection group.
The first tier can reconstruct up to 16 failed HBM pages per group, while the second tier can restore one failed first-tier parity page.
If retained state cannot be reconstructed, \sys falls back to checkpoint recovery.

\subsection{Retained State Lifecycle}
Our design separates memory ownership from data access so that the state manager can retain state without touching potentially faulty pages.
Trainers initialize and update their model state, compute parity pages, and reconstruct failed pages.
RMP allocates memory, maintains allocation and failure metadata, and remaps unavailable virtual pages to healthy memory after failure localization.

\mypara{State allocation.} When training begins, a trainer requests each retained state from RMP by a stable name and size.
RMP allocates HBM pages for each retained state and returns handles to the trainer.
For model state, it additionally allocates the associated first- and second-tier parity pages in host memory.
For reusable process state, RMP retains the HBM allocation without parity protection because its contents are temporary intermediate values that need not survive a failure.
Failed pages are remapped to healthy memory without restoring their contents.
A replacement trainer uses the same name to reattach to the retained allocation.

\mypara{Failure localization.} After a \hof, RMP identifies unavailable pages before the promoted trainer accesses retained state.
For host memory, it scans coarse-grained regions with the \textsf{madvise} system call~\cite{linux_madvise} and recursively bisects any region that returns \textsf{EHWPOISON}, indicating that it contains a failed page.
For HBM, RMP checks each page's mapping with a device query API (\eg \textsf{cuPointerGetAttribute}~\cite{cuda_driver_api}).
RMP marks the failed pages and remaps them to healthy memory.
Remapping restores usable storage; protected contents still require reconstruction from parity.

\mypara{Page reconstruction.} The promoted trainer receives the retained-state handles and the failed-page marks.
For protected state, second-tier parity repairs missing first-tier parity pages, and first-tier parity repairs failed HBM pages.
The trainer must use data and parity from a consistent update boundary for reconstruction.
RMU coordinates reconstruction and recovery from interrupted model updates (\autoref{sec:rmu}).

\section{Crash-Consistent Model Update}\label{sec:rmu}
Retaining model state and protecting its pages does not ensure that it remains consistent after a failure.
A \hof can interrupt in-place updates to the model state and its parity, leaving them inconsistent even when no pages are lost.
We design the \emph{Resumable Model Updater} (RMU) to execute these updates within each trainer and make them crash-consistent with low steady-state and GPU-memory overhead.

\begin{figure}[t]
  \centering
  \includegraphics[width=\columnwidth]{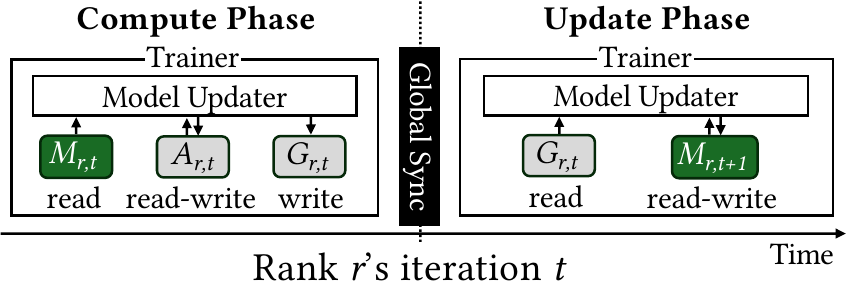}
  \caption{Read, write, and read-write accesses to model state ($M$), activation buffers ($A$), and gradients ($G$) during the compute and update phases of rank $r$'s iteration $t$.}
  \Description{}
  \label{fig:llm_iter}
\end{figure}

\subsection{The Crash-Consistency Problem}
An LLM training iteration consists of a compute phase followed by an update phase (\autoref{fig:llm_iter}).
During the compute phase of iteration~$t$, rank~$r$ computes gradients $G_{r,t}$ from its input batch, using activation buffers $A_{r,t}$ to hold intermediate values for backpropagation.
The model state $M_{r,t}$ remains read-only throughout this phase.
If a \hof interrupts computation, the partial activations and gradients can therefore be discarded and recomputed from the unchanged $M_{r,t}$.
After computing gradients, ranks synchronize to obtain any global gradient statistics needed for the update (\eg the global gradient norm used for clipping~\cite{pascanu2013difficulty}).

During the update phase, the gradients $G_{r,t}$ are read-only, while the model state becomes read-write.
The model updater applies the optimizer (\eg Adam~\cite{adam}) to change the model state in place from $M_{r,t}$ to $M_{r,t+1}$ and refreshes the corresponding host-resident parity.
A \hof can interrupt either operation, leaving the model state partially updated or its parity inconsistent with it.
Recovery must restore a consistent model state and matching parity at each rank and bring all ranks to the same iteration boundary.

\begin{table}[t]
  \centering
  \footnotesize
  \setlength{\tabcolsep}{3pt}
  \caption{Average fault-free iteration time over 30 iterations on \olmoe~(\autoref{tab:workloads}), split into compute and update phases. Reference is the model updater without crash-consistency mechanisms; Sync and Async Snapshotting are the two strawman solutions.}\label{tab:updater-overhead}
  \begin{tabularx}{\columnwidth}{@{}l*{3}{>{\centering\arraybackslash}X}@{}}
    \toprule
    \textbf{Model Updater} & \textbf{Compute (ms)} & \textbf{Update (ms)} & \textbf{Total (ms)} \\
    \midrule
    Reference & 12566 & 47 & 12613 \\
    Sync Snapshotting & 13424 & 82 & 13506 (+7.1\%) \\
    Async Snapshotting & 13372 & 82 & 13454 (+6.7\%) \\
    \rowcolor{gray!12} \textbf{\rmu} & \textbf{12572} & \textbf{111} & \textbf{12683 (+0.6\%)} \\
    \bottomrule
  \end{tabularx}
\end{table}

\mypara{Strawman solution.} A straightforward approach is to snapshot the model state $M_{r,t}$ and its parity into separate host-memory buffers before the update phase.
Each rank completes its snapshot before the global synchronization preceding the update phase, ensuring that no rank starts updating until all snapshots are ready.
After a failure during the update, each rank restores its own pre-update state $M_{r,t}$, and all ranks retry the interrupted iteration.
Once the update completes, each rank encodes new parity for $M_{r,t+1}$ and copies it to the host-resident parity pages.

Preserving a full snapshot, however, imposes substantial steady-state overhead.
The GPU-to-host copies increase iteration time by 7.1\% (\autoref{tab:updater-overhead}).
Asynchronous snapshotting overlaps copies with computation, but resource contention still increases iteration time by 6.7\%.

\subsection{Chunk-Level Transactional Updates}
RMU reduces the cost of crash consistency by preserving progress at chunk boundaries.

\mypara{Key observations.} Two properties of model updates and erasure coding enable this design.
First, modern LLM optimizers update model state at fine granularity.
Adam~\cite{adam} operates element-wise, while Muon~\cite{muon_scalable}, adopted in recent LLM training~\cite{kimilinear}, operates layer-wise.
This structure lets us divide updates into independently committable chunks.
Second, linear erasure codes support incremental parity updates.
We exploit this property of RS coding to update a protection group's parity from each chunk's delta, avoiding recomputation over the entire group.

\begin{figure}[t]
  \centering
  \includegraphics[width=\columnwidth]{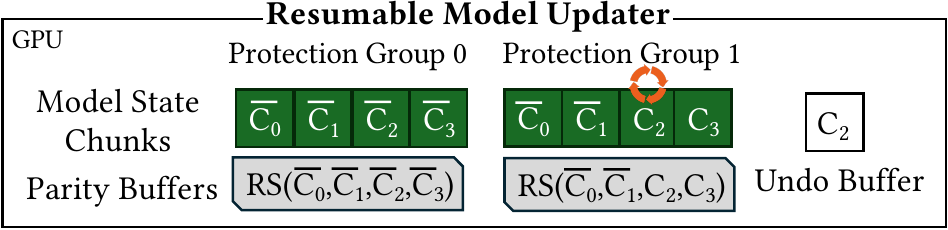}
  \caption{RMU chunk-level transactions. Bars mark committed chunks; $RS(\cdot)$ denotes Reed--Solomon encoding. The undo buffer preserves $C_2$ while its update is in progress.}
  \Description{}
  \label{fig:rmu}
\end{figure}

\mypara{Our approach.}
Building on these observations, we devise \emph{chunk-level transactions} that commit the model state and its parity together.
As shown in~\autoref{fig:rmu}, RMU partitions each RMP protection group into chunks and updates them one at a time.
An \textit{undo buffer} preserves the current chunk's pre-update value, while a \textit{parity buffer} holds parity for the protection group at its latest committed chunk boundary.

In the example, Protection Group~0 has finished updating.
In Protection Group~1, chunks $\overline{C}_0$ and $\overline{C}_1$ are committed, $C_2$ is being updated, and $C_3$ remains unmodified.
The undo buffer holds $C_2$'s pre-update value.
If a \hof interrupts $C_2$'s update, the parity buffer encodes the last committed state, $RS(\overline{C}_0,\overline{C}_1,C_2,C_3)$, whose data remain in the protection group or undo buffer.
Within RMP's configured tolerance, RMU can reconstruct failed pages from this data and parity, restore $C_2$ from the undo buffer, and resume the update.

\mypara{Compute and update phases.} \autoref{alg:leto-update} gives the complete iteration with RMU.
The compute phase begins with the regular forward and backward passes to compute gradients (line~1).
RMU then protects each group's gradients with parity (lines~2--3) before the rank enters the global synchronization (line~4).
This ordering ensures all ranks' gradients are recoverable before any model state changes, preserving the inputs needed to finish an interrupted update.

After synchronization, RMU begins the update phase.
For each protection group, it copies the host-resident parity pages into a temporary GPU parity buffer (line~6).
For each chunk, it preserves the pre-update model state in the undo buffer (line~8) and performs the in-place update (line~9).
It then computes updated parity in a scratch buffer using the chunk delta (line~10).
Committing the chunk makes this scratch parity authoritative and swaps the parity and scratch buffers (line~11).
Once all chunks in the group are committed, RMU copies the resulting parity back to host memory (line~12) and refreshes its second-tier parity.

RMU records its progress in a small host-memory marker that RMP allocates with parity protection.
Lightweight CUDA kernels update the marker in the same stream after each RMU GPU operation.
A promoted trainer uses this marker to identify the last completed operation and the authoritative parity buffer, allowing it to resume the interrupted update correctly.
All ranks retain their committed progress and finish the remaining chunks to complete the same update.

\begin{algorithm}[t]
  \footnotesize
  \DontPrintSemicolon

  \newcommand{\RSenc}[2][]{\FuncSty{RS}_{#1}(#2)}

  \SetKwFunction{Update}{Update}
  \SetKwFunction{HtoD}{Host-to-GPU}
  \SetKwFunction{DtoH}{GPU-to-Host}
  \SetKwFunction{Commit}{Commit}
  \SetKwFunction{Swap}{swap}

  \caption{One training iteration with \rmu}
  \label{alg:leto-update}

  \PlainInput{For $0 \le i < n$, protection group $P_i$ holds
  $k$ chunks $C_j$ with gradients $G_j$, $0 \le j < k$,
  host parity pages $H_i$, parity buffer $Q_i$, and gradient parity
  buffer $R_i$; scratch parity buffer $Q'$; undo buffer $U$.}

Compute gradients\;

  \ForEach{protection group $P_i$}{
    $R_i \gets \RSenc{G_0,\ldots,G_{k-1}}$
    \tcp*[l]{Protect gradients}
  }

Global synchronization\;

  \ForEach{protection group $P_i$}{
    $Q_i \gets \HtoD{$H_i$}$\;

    \ForEach{chunk $C_j$ in $P_i$}{
      $U \gets C_j$
      \tcp*[l]{Preserve pre-update chunk}

      $C_j \gets \Update{$C_j, G_j$}$
      \tcp*[l]{In-place chunk update}

      $Q' \gets Q_i \oplus \RSenc[j]{U \oplus C_j}$
      \tcp*[l]{Parity update}

\Commit{$C_j, Q'$} and \Swap{$Q_i, Q'$}\; }

    $H_i \gets \DtoH{$Q_i$}$\;
  }
\end{algorithm}

\mypara{Minimizing memory and transfer overhead.} Activation buffers and RMU's temporary recovery buffers are needed at different times, allowing them to share GPU memory.
RMU allocates this memory through RMP without separate parity protection because chunk-level commits allow recovery to the last committed chunk boundary after page failures.
During the compute phase, RMU lends this memory to the trainer for activations.
Once backward propagation finishes, it reuses the memory for its gradient-parity, undo, and parity buffers.
To reduce transfer overhead, RMU also pipelines protection groups, overlapping the next group's host-to-GPU transfer and the previous group's GPU-to-host transfer with the current group's computation.
These techniques enable crash-consistent updates without increasing peak HBM usage, with only 0.6\% fault-free iteration overhead (\autoref{tab:updater-overhead}).

\section{OOM-Free Preinitialization}\label{sec:mpreinit}

RMP and RMU keep retained state recoverable and consistent, but the process state prebuilt by a shadow trainer consumes GPU memory alongside active training.
We design the \textit{preinitialization controller} to reclaim this state when active training needs the memory.

\begin{figure}[t]
  \centering
  \includegraphics[width=\columnwidth]{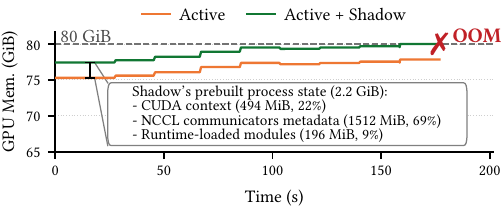}
  \caption{GPU memory usage on \olmoe with global batch size 182. The shadow trainer's prebuilt state causes an OOM in active training after 177\,s.}
  \Description{}
  \label{fig:preinit-oom}
\end{figure}

\mypara{Dynamic memory pressure.}
A shadow trainer may initially fit in GPU memory but later cause an OOM as active memory demand varies, particularly in MoE models where uneven expert routing changes per-GPU token loads~\cite{zhou2022expertchoice,megatron_moe_docs}.
\autoref{fig:preinit-oom} illustrates this effect on \olmoe.
The active trainer alone remains below the 80~GiB GPU capacity, but its shadow adds 2.2~GiB of process state, comprising a 494~MiB CUDA context, 1,512~MiB of NCCL communicator metadata, and 196~MiB of runtime-loaded modules (\eg cuBLAS~\cite{cublas}).
The active trainer then crashes with an OOM error.

\mypara{Our approach.}
Prebuilt process state accelerates recovery but can be discarded without losing training progress.
We exploit this property by integrating reclamation into the active trainer's memory allocation path.
When a request cannot be satisfied, the allocator (\eg PyTorch's \textsf{CUDACachingAllocator}~\cite{cuda_semantics}) invokes the preinitialization controller before reporting an OOM.
The controller synchronously terminates the shadow trainer and waits for its GPU memory to be released.
The allocator then retries the request, transparently avoiding an OOM caused by the shadow.

To avoid repeatedly rebuilding and reclaiming the same state, the controller adjusts how far the shadow preinitializes.
It supports three \textit{preinitialization levels}: (L1) before creating GPU state, (L2) after GPU runtime initialization but before NCCL communicator creation, and (L3) fully preinitialized.
L2 omits NCCL metadata, which accounts for 69\% of the shadow's footprint in~\autoref{fig:preinit-oom}.
After reclamation, the controller relaunches the shadow at the most advanced level that fits the remaining memory headroom.
Further reclamations can only lower this level until the next HOF recovery, when the shadow completes any deferred initialization.
After recovery, the controller uses the observed peak active-training memory to raise the new shadow's level when sufficient headroom is expected, reducing future recovery time.

\section{Implementation}\label{sec:impl}

We implemented \sys{} on Megatron-Core v0.13.1 \cite{megatron-v0-13-1}, PyTorch 2.8.0 \cite{pytorch-2-8-0}, and NCCL 2.27.3 \cite{nccl-2-27-3} in approximately \pythonloc{} lines of Python and \cpploc{} lines of C++/CUDA.

\mypara{Retained state allocation.}
We extend Megatron's model-state tensor allocation path to determine tensor shapes and sizes without allocating GPU memory, using PyTorch's meta device \cite{pytorch-meta-device}.
Based on this metadata, trainers request allocation from RMP over Unix domain sockets.
RMP allocates GPU memory using the CUDA virtual-memory API~\cite{cumemcreate} and host memory using \textsf{memfd\_create}, and transfers the resulting POSIX file descriptors in batches using \textsf{SCM\_RIGHTS} \cite{scm_rights}.
Trainers then import the shared memory and expose it as PyTorch tensors through zero-copy DLPack views \cite{dlpack}.
We also extend NCCL's buffer-allocation path so communicator buffers can be allocated from RMP.

\mypara{Erasure coding computation.}
RMU's Reed--Solomon operations for parity-page encoding and erased-page reconstruction are implemented as CUDA kernels over the finite field $\mathrm{GF}(2^{32})$, using carry-less multiply instructions (\textsf{clmad}~\cite{nvidia_ptx_isa}). Tiled asynchronous memory movement and lazy finite-field reduction overlap data movement with computation and maximize \textsf{clmad} throughput.

\mypara{PyTorch memory allocator integration.}
We implement a custom GPU memory pool in the Megatron runtime and register it with PyTorch's \textsf{CUDACachingAllocator} \cite{cuda_semantics}. The pool reserves temporary buffers required by RMU while allowing the allocator to use this memory for activation buffers during the compute phase. We also add an OOM callback to \textsf{CUDACachingAllocator}; before the allocator reports an OOM, the callback coordinates with the preinitialization controller to reclaim the shadow trainer's GPU memory, after which the allocator retries the failed allocation.

\section{Evaluation}\label{sec:eval}

\subsection{Experimental Setup and Methodology}\label{subsec:setup}

\mypara{Testbed.}
Our AWS cluster has 9 EC2 p4d.24xlarge nodes, each with 8 NVIDIA A100 40\,GB GPUs, 96 vCPUs, 1,152\,GiB host memory, and four 100\,Gbps EFA interfaces (NCCL via aws-ofi-nccl~\cite{aws_efa_nccl}).
Each node in our 3-node private cluster has 2 NVIDIA A100 80\,GB PCIe GPUs, two Intel Xeon Silver 4314 CPUs (32 physical cores), 256\,GiB host memory, and a 100\,Gbps RoCE NIC with GPUDirect RDMA.
Both use 600\,GB/s intra-node NVLink, Linux 6.8.0, NVIDIA driver 580.173.02, and the NGC PyTorch 25.05 container (Python 3.12.3, CUDA 12.9).
Spare-based recovery reserves one warm spare per cluster; the replaced node becomes the next spare.

\begin{table}[t]
  \centering
  \footnotesize
  \caption{Models and parallelism configurations used in our evaluation. DP, TP, PP, and EP denote the degrees of data, tensor, pipeline~\cite{megatron}, and expert~\cite{deepspeed_moe} parallelism, respectively. \emph{Opt.\ states sharded} indicates whether optimizer states are sharded across data-parallel ranks~\cite{zero, fsdp}.}\label{tab:workloads}
  \setlength{\tabcolsep}{2.5pt}
  \begin{tabular*}{\columnwidth}{@{\extracolsep{\fill}}lccccccc@{}}
    \toprule
    \multirow{2}{*}{\textbf{Model}} & \multirow{2}{*}{\makecell{\textbf{Total}\\\textbf{params}}} & \multirow{2}{*}{\makecell{\textbf{Active}\\\textbf{params}}} & \multicolumn{4}{c}{\textbf{Parallelism}} & \multirow{2}{*}{\makecell{\textbf{Opt.\ states}\\\textbf{sharded}}} \\
    \cmidrule(lr){4-7}
    & & & \textbf{DP} & \textbf{TP} & \textbf{PP} & \textbf{EP} & \\
    \midrule
    \qwen~\cite{qwen3} & 4.0B & 4.0B & 2 & 2 & 1 & 1 & \xmark \\
    \olmoe~\cite{olmoe} & 6.9B & 1.3B & 2 & 2 & 1 & 4 & \cmark \\
    \dsmoe~\cite{deepseekmoe} & 16.4B & 2.8B & 8 & 8 & 1 & 16 & \xmark \\
    \gptmoe~\cite{gptoss} & 20.9B & 3.6B & 8 & 8 & 1 & 16 & \cmark \\
    \kimimoe~\cite{kimilinear} & 49.1B & 3.1B & 4 & 4 & 4 & 16 & \cmark \\
    \bottomrule
  \end{tabular*}
\end{table}

\mypara{Workloads.}
We evaluate five dense and MoE LLMs with varied sizes and parallelism strategies (\autoref{tab:workloads}).
We train \qwen and \olmoe on the private cluster with sequence length 512, and the remaining models on AWS with sequence length 2048.
Unless specified otherwise, we use a global batch size of 128; \autoref{subsec:appendix-config} lists other defaults.

\mypara{Baselines.}
We compare \sys against three in-memory checkpointing systems, \gemini~\cite{gemini}, \moevement~\cite{moevement}, and \jitcpkt~\cite{jitckpt}.
We run \moevement only on MoE models, for which its sparse checkpointing is designed.
\jitcpkt requires replicated optimizer state for recovery, so we evaluate it on \qwen and \dsmoe, whose optimizer states are unsharded across DP ranks.
Each baseline also has a ``\textsf{+}'' variant with spare-based preinitialization (\eg \geminiplus; \autoref{fig:failure_handling_by_inmem_preinit}).

\mypara{Recovery policy.}
The three baselines recover in place on a \hof and use spares on an \ff; their ``\textsf{+}'' variants use spares in both cases.
\sys recovers in place on a \hof and adopts \geminiplus for \ffs on \olmoe, \gptmoe, and \kimimoe and \jitcpktplus on \qwen and \dsmoe.

\mypara{Checkpointing interval.}
The \emph{checkpoint interval} is the number of iterations over which the full model state is checkpointed (\eg \moevement's window~\cite{moevement}, and 1 for \jitcpkt, which relies on redundancy in the model state).
We select checkpoint intervals per workload, MTTF, and system (\autoref{subsec:appendix-intervals}).
When \sys uses \geminiplus as its underlying checkpointing system, only \ffs require rollback to a checkpoint.
Given an overall MTTF $M$ and \hof ratio $h$, the \ff rate is $(1-h)/M$, yielding an \ff MTTF of $M/(1-h)$.
We substitute this \ff MTTF into Young's formula~\cite{optcheckpoint} to estimate the optimal checkpoint interval and round it to a whole number of iterations.
At $h=1$, the formula gives no finite optimal interval, so we use the largest interval tested in each setting.

\mypara{Failure injection.}
Each run follows a predefined schedule of injection times, target ranks, and failure classes (\hof or \ff), which determine recovery placement.
We send SIGKILL to the target rank; a monitor detects termination in under 1\,s in our runs and initiates recovery.

\mypara{Metrics.}
We define ETTR and recovery time as follows.

\noindent\textbf{\em ETTR (\%):}
The fraction of wall-clock time spent on productive training (\autoref{subsec:cost}), measured as $\mathrm{ETTR}=T_{\mathrm{ref}}(i)/T_{\mathrm{eval}}(i)$ for the evaluated run's last completed iteration $i$.
Both times span the first iteration's start through iteration $i$'s completion; the reference run is otherwise identical but has no fault-tolerance mechanisms or injected faults.

\noindent\textbf{\em Recovery time (s):}
The time from failure injection until the job recovers its pre-failure training progress, comprising initialization and recomputation.
For a failure at $t_{\mathrm{fault}}$ interrupting iteration $i+1$, let $t_j$ denote when training resumes, $t_{i+1}$ when that iteration completes again, and $t_{\mathrm{remain}}$ its fault-free remaining time at failure.
We measure initialization as $I=t_j-t_{\mathrm{fault}}$, recomputation as $R=t_{i+1}-t_j-t_{\mathrm{remain}}$, and recovery as $I+R$.

\subsection{End-to-End Performance}\label{subsec:e2e-perf}

\begin{figure}[t]
  \centering
  \begin{subfigure}{\columnwidth}
    \centering
    \includegraphics[width=\columnwidth]{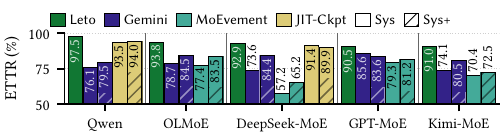}
    \caption{ETTR.}
    \label{fig:overall-ettr}
  \end{subfigure}
  \begin{subfigure}{\columnwidth}
    \centering
    \includegraphics[width=\columnwidth]{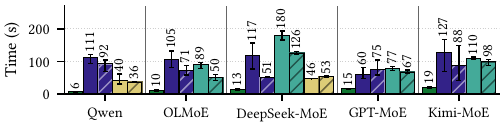}
    \caption{Mean recovery time with min--max error bars.}
    \label{fig:overall-recovery}
  \end{subfigure}
  \caption{Performance with one \hof every 10 minutes. Hatched bars denote spare-based preinitialization (``\textsf{+}''); missing bars indicate unsupported workloads.}
  \Description{}
  \label{fig:overall-performance}
\end{figure}
\mypara{Overall performance.}
We run all applicable systems on the five workloads for 30 minutes, injecting one \hof every 10 minutes (\autoref{fig:overall-performance}).
\sys uses the same checkpoint intervals as \gemini and \geminiplus for \ff recovery on \olmoe, \gptmoe, and \kimimoe.
We examine the impact of longer checkpoint intervals for \sys below.
\sys sustains 91--98\% ETTR, exceeding each workload's best baseline by 1.5--10.4 percentage points.
\sys recovers in 6--19\,s, 3.6--6.5$\times$ faster than the fastest baseline for each workload (\autoref{fig:overall-recovery}).
Baselines replay from checkpoints; \sys reuses the live model state, repeating at most the interrupted iteration.

\begin{figure}[t]
  \centering
  \includegraphics[width=\columnwidth]{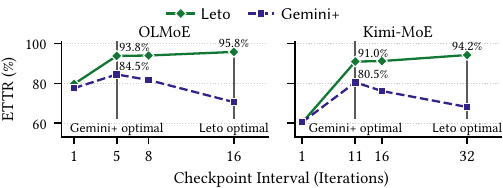}
  \caption{ETTR of \sys and \geminiplus on \olmoe and \kimimoe with one \hof every 10 minutes, varying the checkpoint interval; vertical lines mark each system's ETTR-maximizing interval.}
  \Description{}
  \label{fig:checkpoint-interval}
\end{figure}

\mypara{Impact of the checkpoint interval.}
\autoref{fig:checkpoint-interval} examines sensitivity to the checkpoint interval on \olmoe (1, 5, 8, and 16 iterations) and \kimimoe (1, 11, 16, and 32) with only \hofs.
\geminiplus peaks at intervals of 5 and 11, reaching 84.5\% and 80.5\% ETTR, respectively; shorter intervals incur more checkpointing overhead, while longer intervals increase recomputation.

\sys retains checkpoints for \ffs, but recovers from \hofs without rollback.
In this HOF-only experiment, longer intervals therefore reduce checkpointing overhead without increasing recomputation.
\sys's ETTR rises from 93.8\% to 95.8\% on \olmoe as the interval grows from 5 to 16, and from 91.0\% to 94.2\% on \kimimoe as it grows from 11 to 32.
At these intervals, \sys exceeds \geminiplus's peak ETTR by 11.3 and 13.7 percentage points, respectively.

\begin{figure}[t]
  \centering
  \includegraphics[width=\columnwidth]{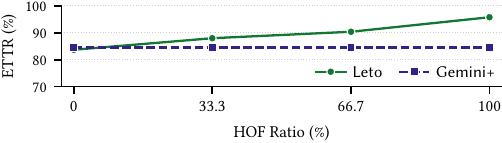}
  \caption{ETTR of \sys and \geminiplus on \olmoe with one failure every 10 minutes, varying the \hof ratio.}
  \Description{}
  \label{fig:hof-ratio}
\end{figure}

\mypara{Impact of the HOF ratio.}
We next vary the \hof ratio from 0\% (all \ffs) to 100\% on \olmoe (\autoref{fig:hof-ratio}), comparing against \geminiplus, its best baseline in \autoref{fig:overall-performance}.
\geminiplus uses its tuned interval of 5; we reuse its measured 84.5\% ETTR across ratios because both failure classes follow the same recovery path.
Using the interval-estimation method in \autoref{subsec:setup}, we set \sys's interval to 5, 6, and 9 iterations at 0\%, 33.3\%, and 66.7\% \hofs; at 100\%, we use the largest tested interval of 16.
Its ETTR rises from 83.7\% to 95.8\%, exceeding \geminiplus by 3.5 percentage points even at a 33.3\% \hof ratio.
With only \ffs, its steady-state overhead costs 0.8 percentage points.

\begin{figure}[t]
  \centering
  \includegraphics[width=\columnwidth]{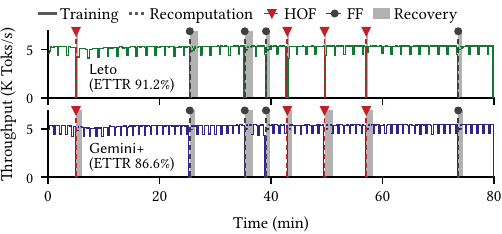}
  \caption{Throughput on \olmoe with failure arrivals from a public trace and a 50\% \hof ratio. Shading marks recovery, dotted segments recomputation, and periodic notches checkpointing.}
  \Description{}
  \label{fig:realistic-faults}
\end{figure}

\mypara{Realistic failures with a public trace.}
To emulate irregular training failures, we schedule injections using eviction times from an 80-minute segment of a public GCP A100 spot-instance trace used in prior work~\cite{bamboo,oobleck,moevement}.
The segment contains eight eviction events, averaging one every 10 minutes.
We inject a training failure at each event, randomly assigning four as \hofs and four as \ffs.

\autoref{fig:realistic-faults} compares \sys and \geminiplus on \olmoe at their optimal checkpoint intervals of 8 and 5, respectively.
\sys achieves 91.2\% ETTR versus 86.6\% for \geminiplus and reduces total recovery time from 7.8 to 4.6 minutes, or from 59 to 34\,s per failure on average.
Its \hof recovery averages 11\,s, while \ff recovery averages 57\,s using \geminiplus.
Fast \hof recovery therefore improves training efficiency even with irregular arrivals and checkpoint rollback for half the failures.

\subsection{Performance Breakdown}\label{subsec:perf_breakdown}

\begin{table}[t]
  \centering
  \footnotesize
  \setlength{\tabcolsep}{3pt}
  \caption{Recovery-time breakdown and ETTR on \kimimoe for \gemini, \geminiplus, and \sys at the indicated checkpoint intervals. Interval 11 maximizes ETTR for \gemini and \geminiplus, while 32 does so for \sys (\autoref{fig:checkpoint-interval}).}\label{tab:recovery-breakdown}
  \begin{tabularx}{\columnwidth}{@{}l*{4}{>{\centering\arraybackslash}X}@{}}
    \toprule
    \textbf{System} & \textbf{Recovery} & \textbf{Init.} & \textbf{Recomp.} & \textbf{ETTR} \\
    \midrule
    \gemini (interval=11) & 127\,s & 59\,s & 68\,s & 74.1\% \\
    \geminiplus (interval=11) & 88\,s & 35\,s & 53\,s & 80.5\% \\
    \sys (interval=11) & 19\,s & 15\,s & 5\,s & 91.0\% \\
    \rowcolor{gray!12} \textbf{\sys (interval=32)} & \textbf{16\,s} & \textbf{13\,s} & \textbf{3\,s} & \textbf{94.2\%} \\
    \bottomrule
  \end{tabularx}
\end{table}

\autoref{tab:recovery-breakdown} separates recovery on \kimimoe into initialization and recomputation to show where \sys saves time.
Spare-based preinitialization reduces initialization from 59\,s to 35\,s (\gemini to \geminiplus).
\rmp and \rmu reduce recomputation from 53\,s to 5\,s.
\rmp retains the live model state to avoid checkpoint replay, while \rmu resumes an interrupted update from its last committed chunk.
In-place preinitialization further reduces initialization from 35\,s to 15\,s by letting the shadow take over with state prebuilt for the same hardware and rank.
At the same checkpoint interval of 11, these mechanisms yield 4.5$\times$ faster recovery than \geminiplus (88\,s to 19\,s) and raise ETTR from 80.5\% to 91.0\%.
Because only \ffs require checkpoint rollback, \sys can further reduce checkpointing overhead by extending its interval to 32, reaching 94.2\% ETTR with 16\,s recovery.

\subsection{Recovery Correctness}\label{subsec:correctness}

\begin{figure}[t]
  \centering
  \includegraphics[width=\columnwidth]{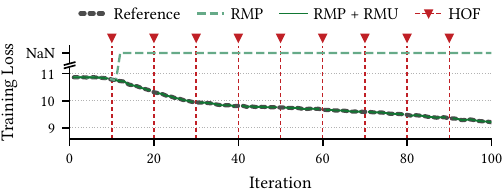}
  \caption{Training loss of \olmoe over 100 iterations for the fault-free reference, \rmp alone, and \rmp with \rmu, with a \hof injected during the update phase every 10 iterations.}
  \Description{}
  \label{fig:recovery-correctness}
\end{figure}

We verify that \sys recovers exactly from \hofs that interrupt the update phase and lose part of the retained state.
We train \olmoe for 100 iterations and, every 10 iterations, kill one rank during its model update, interrupting the optimizer kernel, the parity-encoding kernel, or the parity copy in turn; we instrument the kernels with NVBit~\cite{nvbit} to abort mid-execution and fill the copy's destination with random values.
After each failure, we emulate uncorrectable HBM errors by unmapping retained \rmp pages with \texttt{cuMemUnmap}, 16 per protection group, the maximum its parity tolerates, and a host-memory error on one parity page by poisoning it with \texttt{madvise(MADV\_HWPOISON)}~\cite{linux_hwpoison}, following prior work~\cite{pangolin}.

\autoref{fig:recovery-correctness} shows the training loss.
With \rmp alone, the parity is inconsistent with the partially updated model state, so the state reconstructed after the first \hof is wrong: the loss diverges from the reference at the next iteration and becomes NaN thereafter.
With \rmp and \rmu, \sys recovers exactly from every injected \hof, reproducing the fault-free reference loss.
Per recovery, \rmp localizes failures in 36\,ms and \rmu reconstructs pages in 140\,ms.

\subsection{Recovery under Memory Pressure}\label{subsec:memory-pressure}
\begin{figure}[t]
  \centering
  \includegraphics[width=\columnwidth]{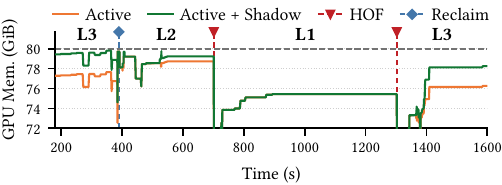}
\caption{GPU memory usage of \sys on \olmoe under memory pressure. The gap between the curves is the shadow's footprint. Levels denote full preinitialization (L3), GPU runtime only (L2), and no GPU state (L1).}
  \Description{}
  \label{fig:memory-pressure}
\end{figure}

We evaluate whether the preinitialization controller (\autoref{sec:mpreinit}) prevents OOMs from shadow memory use while preserving recovery benefits.
The \emph{preinitialization level} specifies how far the shadow trainer initializes before a failure.
We train \olmoe at global batch size 182 and inject one \hof every 10 minutes for an hour. Without runtime reclamation, the fully preinitialized shadow causes an OOM (\autoref{fig:preinit-oom}).

As \autoref{fig:memory-pressure} shows for the first three \hofs of the run, the controller adapts the shadow to available memory.
When memory pressure rises, it reclaims the 2.2\,GiB L3 shadow and relaunches it at L2.
If L2 no longer fits, the controller lowers the shadow to L1; it restores L3 once pressure subsides.
Despite these transitions, every \hof recovers through a shadow, taking 26--27\,s from L3, 33--34\,s from L2, and 52\,s from L1. Over the hour, \sys achieves 91.8\% ETTR, 4.1 percentage points above \geminiplus: reclamation gives up only as much preinitialization as memory pressure demands, and \hof recovery never rolls back to a checkpoint.

\subsection{Large-Scale Simulation}

We evaluate scalability beyond our testbed using a simulator that extrapolates single-stage profiles to predict iteration time, recovery time, and ETTR under a given failure schedule.
We validate the simulator against training performance measured on our GPU testbed, with ETTR prediction errors of 3.1\% and 0.7\% for \geminiplus on \dsmoe and \kimimoe, respectively.
We will open-source the simulator.
See \autoref{sec:appendix-sim} for details.

We increase model and cluster size together, simulating \dsvtwo~\cite{deepseek_v2} (236B total, 21B active parameters) on 1,024 GPUs, \dsvthree~\cite{deepseek_v3} (671B total, 37B active) on 16,384 GPUs, and \dsvfour~\cite{deepseek_v4} (1.6T total, 49B active) on 131,072 GPUs.
For each configuration, we set the MTTF to 10 minutes and vary the \hof ratio over 0\%, 50\%, and 100\%.

\begin{figure}[t]
  \centering
  \includegraphics[width=\columnwidth]{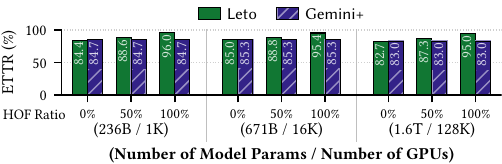}
  \caption{Simulated ETTR of \sys and \geminiplus with a 10-minute MTTF, varying the \hof ratio across model and cluster sizes.}
  \Description{}
  \label{fig:scalability}
\end{figure}

As \autoref{fig:scalability} shows, \sys's advantage grows with the \hof ratio at every scale.
With 100\% \hofs, \sys sustains 95--96\% ETTR, compared with 83.0--85.3\% for \geminiplus.
Even at 50\% \hofs, \sys improves ETTR by 3.5--4.3 percentage points; with no \hofs, it trails by just 0.3 percentage points.
These results suggest that state reuse remains beneficial at scale even when \hofs account for only half of all failures.
See \autoref{sec:appendix-ettr-accuracy} for results at 30-minute and 1-hour MTTFs.

\section{Related Work}\label{sec:related}
\mypara{Checkpointing systems.}
Gemini~\cite{gemini} keeps checkpoints in remote host memory, while MoEvement~\cite{moevement} exploits MoE sparsity for lightweight checkpointing and localized recovery.
JIT-Checkpointing~\cite{jitckpt} requires replicated optimizer state, precluding its use with optimizer-state sharding in ZeRO and FSDP~\cite{zero,fsdp}.
CheckFreq~\cite{checkfreq} adapts checkpoint frequency and overlaps checkpointing with training, while PCcheck~\cite{pccheck} pipelines transfers to persistent storage and supports concurrent checkpoints.
MoC-System~\cite{moc} checkpoints a subset of experts; ByteCheckpoint~\cite{bytecheckpoint} and Universal Checkpointing~\cite{ucp} support recovery under different parallelism configurations.
Checkmate~\cite{checkmate} maintains per-iteration recovery state using dedicated CPU nodes and programmable switches.
\sys{} instead reuses working state for in-place \hof recovery without loading checkpoints.

\mypara{Runtime reconfiguration and redundant computation.}
Runtime reconfiguration~\cite{varuna,oobleck,parcae,recycle} tolerates resource loss by adapting training to fewer nodes.
\sys{} instead exploits surviving hardware to preserve the job's scale and avoid reconfiguration costs~\cite{trainmover}.
Bamboo~\cite{bamboo} avoids checkpoint rollback through redundant computation for neighboring pipeline stages, but incurs steady-state overhead~\cite{oobleck}.
\sys{} requires no redundant training computation.

\mypara{Erasure protection for memory-resident state.}
Prior systems use checksums and erasure coding to protect persistent-memory and memory-resident data~\cite{novafortis,pangolin,pavise,carbink}, some maintaining consistent redundancy across transactions~\cite{pangolin,pavise}.
\sys{} protects GPU-resident model state without storing persistent parity in HBM.
ECRM~\cite{ecrm} erasure-codes embedding tables with sparse updates to limit network overhead, but replicates dense parameters.
\sys{} keeps model state and parity consistent across dense optimizer updates.

\section{Conclusion}
\label{sec:conclusion}

We presented \sys, which decouples training-state and process lifetimes to recover from \hofs through state reuse and advance preparation on surviving hardware.
Resilient state retention, crash-consistent updates, and reclaimable preinitialization deliver 3.6--6.5$\times$ faster recovery than the best-performing checkpointing baselines and up to 13.7 percentage points more productive training time.
\sys opens new opportunities for accelerator runtimes to preserve application progress across process failures.

\bibliographystyle{ACM-Reference-Format}
\bibliography{references}

\clearpage
\appendix
\section{Evaluation Details}\label{sec:appendix-eval}

\subsection{Additional Default Training Configuration}\label{subsec:appendix-config}
We train with BF16 mixed precision~\cite{mixed_precision, bf16} and apply activation checkpointing at the input to every layer~\cite{chen2016training}.
Microbatch sizes are 4 for the pipeline-parallel model and 16 for the others. For the pipeline-parallel model (\ie \kimimoe), we use interleaved one-forward-one-backward (1F1B) scheduling with two model chunks per pipeline-parallel rank~\cite{megatron}. For the other models, the global batch size of 128 is divided evenly across the DP ranks. We use Adam~\cite{adam} with $\beta_1{=}0.9$, $\beta_2{=}0.95$, and weight decay $0.1$. The learning rate warms up over the first $10\%$ of iterations to a peak of $2.2\times10^{-4}$, then follows a cosine schedule to a floor of $2.2\times10^{-5}$ (one tenth of the peak) by the end of the run.
Gradients are clipped to a global norm of $1.0$, weights are initialized from $\mathcal{N}(0,\,0.006^{2})$, and the attention softmax is computed in FP32.
All runs train on FineWeb~\cite{fineweb}.
We enable the deterministic modes of PyTorch and CUDA libraries~\cite{pytorch-reproducibility, cublas-reproducibility} so that training is bit-reproducible and recovered runs can be compared exactly against the fault-free reference.

\begin{table}[H]
  \centering
  \footnotesize
  \caption{Tuned checkpoint interval $I$ and sparse window $W$, in iterations, per workload and MTTF. Dashes indicate that \moevement is not evaluated on the dense workload, \qwen.}
  \label{tab:intervals}
  \setlength{\tabcolsep}{2.5pt}
  \begin{tabular*}{\columnwidth}{@{\extracolsep{\fill}}lcccccc@{}}
    \toprule
    \multirow{2}{*}{\textbf{Workload}} & \multicolumn{3}{c}{$I$ (\gemini, \geminiplus)}
    & \multicolumn{3}{c}{$W$ (\moevement, \moevementplus)} \\
    \cmidrule(lr){2-4}\cmidrule(lr){5-7}
    & \textbf{10\,min} & \textbf{30\,min} & \textbf{1\,hr} & \textbf{10\,min} & \textbf{30\,min} & \textbf{1\,hr} \\
    \midrule
    \qwen     & 22 & 37 & 53 & -- & -- & -- \\
    \olmoe    &  5 &  9 & 12 & 2  & 4  & 4  \\
    \dsmoe    & 15 & 25 & 36 & 8  & 8  & 8  \\
    \gptmoe   & 12 & 21 & 29 & 4  & 8  & 8  \\
    \kimimoe  & 11 & 19 & 27 & 4  & 8  & 8  \\
    \bottomrule
  \end{tabular*}
\end{table}

\subsection{Checkpoint Intervals}\label{subsec:appendix-intervals}
  For interval-based checkpointing systems, we measure the normalized per-checkpoint cost $p$ as the excess time of an iteration that commits a checkpoint over one that does not, divided by the uncheckpointed iteration time.
  We apply Young's formula~\cite{optcheckpoint} to estimate the checkpoint interval in iterations as $I^\star = \sqrt{2p\,\mathrm{MTTF}/t_{\mathrm{iter}}}$, where $t_{\mathrm{iter}}$ is the uncheckpointed iteration time.
  We use the same checkpoint-cost measurements for \geminiplus and \gemini, since they differ in recovery placement rather than checkpoint cost.

  For window-based checkpointing systems, we select the window with the lowest estimated cost among those that fit in the pinned-memory pool, using a calibrated sparse-window model.
  \jitcpkt and \jitcpktplus commit every iteration by construction ($I=1$), on the two workloads whose optimizer states are replicated (\qwen and \dsmoe).

  \sys maintains checkpoints for \ffs, while its \hof recovery does not load a checkpoint.
  In the HOF-ratio experiments, we estimate its checkpoint interval using the mean time between fatal failures, $M/(1-h)$, as described in \autoref{subsec:setup}.
  \autoref{tab:intervals} lists baseline intervals; the evaluation specifies the intervals used in each experiment.

\section{Simulated ETTR at Longer MTTFs}\label{sec:appendix-ettr-accuracy}
\begin{figure}[H]
  \centering
  \begin{subfigure}{\columnwidth}
    \centering
    \includegraphics[width=\columnwidth]{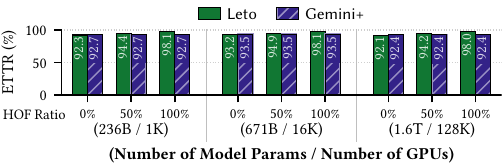}
    \caption{MTTF of 30 minutes.}
    \label{fig:scalability-30m}
  \end{subfigure}
  \begin{subfigure}{\columnwidth}
    \centering
    \includegraphics[width=\columnwidth]{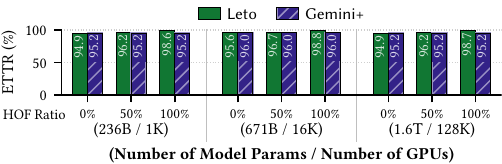}
    \caption{MTTF of 1 hour.}
    \label{fig:scalability-1hr}
  \end{subfigure}
  \caption{Simulated ETTR at longer MTTFs, using the model and cluster configurations in \autoref{fig:scalability}.}
  \Description{}
  \label{fig:scalability-extended}
\end{figure}

We predict ETTR for \sys and \geminiplus at MTTFs of 30 minutes and 1 hour, extending the results in \autoref{fig:scalability}.
As \autoref{fig:scalability-extended} shows, \sys's advantage again grows with the \hof ratio at every scale: it leads \geminiplus by 0.7--1.8 percentage points at 50\% \hofs and by 2.8--5.6 points at 100\%, and trails by 0.3 to 0.4 points with 0\% HOFs due to its steady-state overhead.

\section{Simulation Details}\label{sec:appendix-sim}

This section describes the simulator used to predict ETTR for the large-scale training scenarios in \autoref{fig:scalability}.
\autoref{tab:large_configs} lists the model and parallelism configurations used in these simulations.

\begin{table}[H]
  \centering
  \footnotesize
  \caption{Models and parallelism configurations used in our large-scale simulations. DP, TP, PP, and EP denote the degrees of data, tensor, pipeline~\cite{megatron}, and expert~\cite{deepspeed_moe} parallelism, respectively. \emph{Opt.\ states sharded} indicates whether optimizer states are sharded across data-parallel ranks~\cite{zero, fsdp}.}\label{tab:large_configs}
  \setlength{\tabcolsep}{2.5pt}
  \begin{tabular*}{\columnwidth}{@{\extracolsep{\fill}}lccccccc@{}}
    \toprule
    \multirow{2}{*}{\textbf{Model}} & \multirow{2}{*}{\makecell{\textbf{Total}\\\textbf{params}}} & \multirow{2}{*}{\makecell{\textbf{Active}\\\textbf{params}}} & \multicolumn{4}{c}{\textbf{Parallelism}} & \multirow{2}{*}{\makecell{\textbf{Opt.\ states}\\\textbf{sharded}}} \\
    \cmidrule(lr){4-7}
    & & & \textbf{DP} & \textbf{TP} & \textbf{PP} & \textbf{EP} & \\
    \midrule
    \dsvtwo~\cite{deepseek_v2} & 236B & 21B & 8 & 8 & 16 & 16 & \cmark \\
    \dsvthree~\cite{deepseek_v3} & 671B & 37B & 128 & 8 & 16 & 64 & \cmark \\
    \dsvfour~\cite{deepseek_v4} & 1.6T & 49B & 1024 & 8 & 16 & 64 & \cmark \\
    \bottomrule
  \end{tabular*}
\end{table}

\subsection{Modeling Iteration Time}\label{subsec:appendix-model}
To predict a training system's ETTR, the simulator must accurately model its iteration time ($T_{\mathrm{iter}}$).
We take inspiration from existing simulators for LLM workloads \cite{dooly,vidur,llmservingsim} and use a profile-based approach to simulation.
Given a deployment scenario and model, we profile the training job at the granularity of a single pipeline stage and use the profile to predict the iteration time of the entire job.

We model iteration time as the sum of the pipeline, synchronization, and update times.
These terms correspond to forward and backward propagation under the pipeline schedule, data-parallel gradient synchronization, and the optimizer step, respectively.
$$
T_{\mathrm{iter}}=T_{\mathrm{pipeline}}+T_{\mathrm{sync}}+T_{\mathrm{update}}
$$

\subsubsection{Pipeline Time ($T_{\mathrm{pipeline}}$)} \label{subsec:appendix-pipeline}

$T_{\mathrm{pipeline}}$ covers the scheduled forward and backward passes.
Its profiling unit is the \emph{bracket}, a contiguous chunk of layers within one pipeline stage, traversed forward and backward by a single microbatch.
We measure each bracket in isolation using the deployment's hardware placement, including all intra-layer communication (tensor-parallel collectives and MoE all-to-all) in the timed interval.
Given bracket times, we estimate the pipeline time as
$$T_{\mathrm{pipeline}} = \left(M + \frac{S-1}{v}\right)\cdot\max_s \mathrm{stage}(s)$$

Here, $M$ is the number of microbatches, $S$ is the number of pipeline stages, and $v$ is the virtual-pipeline interleaving degree.
$\mathrm{stage}(s)$ sums the times for stage $s$'s brackets and inter-stage point-to-point transfers, plus the embedding on the first stage or the output head and loss on the last stage.
We estimate transfer times from a fitted send/receive curve at the profiled activation size; the isolated bracket measurements exclude these transfers.
Taking the maximum captures the bottleneck imposed by the slowest concurrent stage.

\subsubsection{Synchronization and Update Time ($T_{\mathrm{sync}}+T_{\mathrm{update}}$)} \label{subsec:appendix-sync-update}

$T_{\mathrm{sync}}$ is the time for data-parallel gradient synchronization, modeled using collective operations that we profile in isolation and fit to a parametric model.
We also profile concurrent collectives to capture the cost of two collectives sharing the same network fabric.

$T_{\mathrm{update}}$ is the optimizer-step time, modeled using machine and network constants so that no per-model sweep is required.
We profile the optimizer step and scalar reduction in isolation and fit their timings to a parametric model.
For sharded optimizers, we also profile and model the all-gather that reconstructs the full parameter tensor for the next forward pass.

\subsubsection{Communication--Computation Overlap}

Overlapping gradient synchronization with backward computation affects $T_{\mathrm{iter}}$, but individual timers do not directly capture the resulting interference.
We therefore measure interference differentially, running the same configuration with overlap disabled and enabled.
The measured saving is $\mathit{saving} = T_{\mathrm{iter}}^{\mathrm{off}} - T_{\mathrm{iter}}^{\mathrm{on}}$, while a lightweight event logger in the enabled run records the exposed communication time, $E$.
Because the two runs differ only in the overlap setting, we decompose the serial synchronization time $T_{\mathrm{sync}}$ into time saved, exposed communication, and residual interference.

$$
T_{\mathrm{sync}} = \mathit{saving} + E + R .
$$

The residual $R$ captures the additional cost of resource contention between the in-flight collective and backward computation.

\mypara{Modeling $E$.}
We replay the framework's own overlap mechanism as a queue, with no fitted parameter.
Backward walks the stage's layers in reverse over a window of length $W$, taken from the bracket profile of \autoref{subsec:appendix-pipeline}.
Megatron packs gradients into fixed-size \emph{buckets} of $\max(40\mathrm{M},\,1\mathrm{M}\cdot d_p)$ parameters, where $d_p$ is the data-parallel degree and $\mathrm{M}$ denotes one million.
A bucket's collective launches once backward propagation produces its last gradient.
Bucket $i$ has a transfer time $w_i$, obtained by evaluating the gradient-collective model of \autoref{subsec:appendix-sync-update} at its byte size and multiplying by the measured concurrency penalty.
We approximate its ready time as $r_i = \rho_i W$, where $\rho_i$ is the fraction of the stage's parameters held by buckets $1,\ldots,i$, assuming backward progress is proportional to the number of parameters processed.
The framework keeps one collective in flight, so buckets use a single serial channel.
Bucket $i$ finishes at $f_i = \max(r_i,\, f_{i-1}) + w_i$, with $f_0=0$.
For $n$ buckets, the communication remaining after the backward window is
$$
E = \max(0,\; f_n - W).
$$

\mypara{Modeling $R$.}
We model $R$ using a calibrated constant for each deployment class with interference, normalizing the measured residual by the quantity that drives it.
For \emph{dense} models, no other traffic shares the fabric during backward computation, so interference arises from contention with computation.
We model this cost as proportional to the communication hidden under computation, $T_{\mathrm{sync}} - E$, giving $c = R / (T_{\mathrm{sync}} - E)$ and arrive at
$$
T_{\mathrm{sync}}^{\mathrm{on}} = E \;+\; c \cdot (T_{\mathrm{sync}} - E).
$$

For \emph{MoE} models whose expert groups span nodes, the gradient collective ($G$) and MoE all-to-all ($A$) contend for the same inter-node fabric.
Two flows sharing a bottleneck finish no later than in series ($G + A$) and no earlier than the longer alone ($\max(G, A)$).
This means that the delay attributable to sharing lies in $[0, \min(G, A)]$.
The calibrated fraction is $f = R / \min(G, A)$, always lies within the range $[0, 1]$ for the profiled deployments.
We therefore model the effective synchronization time as:

$$
T_{\mathrm{sync}}^{\mathrm{on}} = E' \;+\; f \cdot \min(G, A),
$$

where $E'$ is the queue's exposure evaluated against the stretched window $W + f\cdot\min(G,A)$.
MoE deployments whose all-to-all stays within a node use no interference constant.
When overlap is enabled, we substitute $T_{\mathrm{sync}}^{\mathrm{on}}$ for $T_{\mathrm{sync}}$ in the iteration-time model.

\subsection{Iteration Time Prediction Accuracy}\label{subsec:appendix-t-iter}

With the above methods, we predict the iteration time of the models in \autoref{tab:workloads}.
As shown in \autoref{tab:t-iter-accuracy}, the predicted iteration times have a mean absolute percentage error of 2.18\%.

\begin{table}[H]
  \centering
  \footnotesize
  \setlength{\tabcolsep}{2pt}
  \caption{Simulator accuracy on iteration time. Predicted and measured
  $T_{\mathrm{iter}}$ (s) for every workload in \autoref{tab:workloads}, with
  communication--computation overlap enabled. Measurements are the median
  of three independent runs from iteration 11 onward.}\label{tab:t-iter-accuracy}
  \begin{tabular*}{\columnwidth}{@{\extracolsep{\fill}}lccccc@{}}
    \toprule
    & \qwen & \olmoe & \dsmoe & \gptmoe & \kimimoe \\
    \midrule
    Predicted (s) & 3.464 & 11.939  & 5.742 & 5.198 & 9.246 \\
    Measured (s)  & 3.556 & 12.002 & 5.927 & 5.363 & 9.097 \\
    Error (\%)    & $-2.6$ & $-0.5$ & $-3.1$ & $-3.1$ & $+1.6$ \\
    \bottomrule
  \end{tabular*}
\end{table}

\subsection{ETTR Prediction}\label{subsec:appendix-ettr}
To predict ETTR, we model both the reference and evaluated execution times.

The reference run uses no fault-tolerance mechanisms and experiences no failures, so every iteration takes the unmodified iteration time $T_{\mathrm{iter}}$.

\begin{equation}
  T_{\mathrm{ref}}(i) = i\,T_{\mathrm{iter}}.
  \label{eq:tref}
\end{equation}

The evaluated run differs in two ways.
Each iteration takes $T_{\mathrm{eff}}$ rather than $T_{\mathrm{iter}}$, because fault-tolerance mechanisms add overhead even without failures.
Each failure also incurs recovery time $T_{\mathrm{rec}}$.
Failures arrive at rate $1/\mathrm{MTTF}$ in elapsed time, recovery included, so a run lasting $T_{\mathrm{eval}}$ experiences $T_{\mathrm{eval}}/\mathrm{MTTF}$ failures in expectation.
We therefore model the elapsed time as

\begin{equation}
  T_{\mathrm{eval}}(i) = i\,T_{\mathrm{eff}}
  \;+\; \frac{T_{\mathrm{eval}}(i)}{\mathrm{MTTF}}\,T_{\mathrm{rec}}.
  \label{eq:teval-implicit}
\end{equation}

Collecting the terms in $T_{\mathrm{eval}}(i)$ and solving gives

\begin{align}
  T_{\mathrm{eval}}(i) - T_{\mathrm{eval}}(i)\,\frac{T_{\mathrm{rec}}}{\mathrm{MTTF}}
    &= i\,T_{\mathrm{eff}},
  \label{eq:collect} \\[3pt]
  T_{\mathrm{eval}}(i)\left(1 - \frac{T_{\mathrm{rec}}}{\mathrm{MTTF}}\right)
    &= i\,T_{\mathrm{eff}},
  \label{eq:factor} \\[3pt]
  T_{\mathrm{eval}}(i)
    &= \frac{i\,T_{\mathrm{eff}}}{\,1 - T_{\mathrm{rec}}/\mathrm{MTTF}\,}.
  \label{eq:teval}
\end{align}
Dividing \autoref{eq:tref} by \autoref{eq:teval} cancels the iteration count, giving

\begin{equation}
\begin{aligned}
  \mathrm{ETTR}
  &= \frac{i\,T_{\mathrm{iter}}}
         {i\,T_{\mathrm{eff}} \big/ \left(1 - T_{\mathrm{rec}}/\mathrm{MTTF}\right)} \\
  &= \frac{T_{\mathrm{iter}}}{T_{\mathrm{eff}}}
    \left(1 - \frac{T_{\mathrm{rec}}}{\mathrm{MTTF}}\right).
\end{aligned}
  \label{eq:ettr}
\end{equation}

The first factor captures the effect of steady-state overhead; the second captures the effect of recovery time.

\autoref{eq:ettr} holds under four assumptions.
(1) $T_{\mathrm{rec}} < \mathrm{MTTF}$: recovery is short relative to the time between failures, so the second factor stays positive and its linear form is accurate; the ratio reaches $0.3$ at most in our evaluation.
(2) Failures do not overlap: each is fully recovered before the next arrives, so every failure is charged one independent $T_{\mathrm{rec}}$; no run in our evaluation contains a failure inside another's recovery.
(3) Failures arrive uniformly at random in time, which gives the expected rollback of $I/2 + 1$ iterations for the interval-based systems, which overlaps checkpointing with next iteration's computation phase, and half an iteration for \sys.
(4) The run is long enough that the failure count is near its expectation and each recovery costs the same $T_{\mathrm{rec}}$.

We decompose recovery time as
$T_{\mathrm{rec}} = T_{\mathrm{det}} + T_{\mathrm{init}} + L\,T_{\mathrm{eff}}$
where $T_{\mathrm{det}}$ is the failure-detection time, $T_{\mathrm{init}}$ is the time to prepare a replacement trainer, and $L\,T_{\mathrm{eff}}$ is the time to recompute $L$ iterations of lost progress.

\autoref{eq:ettr} applies to both systems, with the following system-specific terms.
\begin{align}
  \text{Gemini+:} && T_{\mathrm{eff}} &= T_{\mathrm{iter}} + T_{\mathrm{ckpt}}/I,
                  && L &= \tfrac{I}{2} + 1,
  \label{eq:gplus} \\[3pt]
  \text{Leto:}    && T_{\mathrm{eff}} &= (1+\omega)\,T_{\mathrm{iter}},
                  && L &= \tfrac{1}{2}.
  \label{eq:leto}
\end{align}

During recovery, \geminiplus restores checkpointed state on a promoted standby and redoes the work since the newest usable checkpoint.
A checkpoint is taken at the boundary of every $I$-th iteration but becomes usable only once its transfer completes, at the following iteration's step, so a uniformly arriving failure loses between $1$ and $I$ completed iterations, $(I+1)/2$ in expectation, plus the elapsed half of the iteration in flight: $L = I/2 + 1$.
\sys loses no completed iteration and only the in-flight half, $L = 1/2$.

\sys keeps a checkpointing interval for \ff recovery, so its steady cost is $T_{\mathrm{ckpt}}/I$ plus the per-iteration overhead of RMU $\omega$. We estimate $\omega$ by profiling the duration of RMU's update phase and divide it by the estimated iteration time.
Recovery promotes the shadow without transferring state; the only lost work is the elapsed portion of the interrupted iteration, half an iteration in expectation.

\end{document}